%% file: main.tex
\documentclass{article}
\usepackage{spconf,amsmath,graphicx}
\usepackage{amsmath}
\usepackage{amssymb}
\usepackage{booktabs}
\usepackage{float}
\usepackage{subcaption}
\usepackage{setspace}
\usepackage{hyperref}
\usepackage{xcolor}
\hypersetup{
    colorlinks=true,    
    linkcolor=red,      
    citecolor=blue,     
    filecolor=magenta,  
    urlcolor=red       
}

\title{ICP-3DGS: SfM-Free 3D Gaussian Splatting \\ for large-scale unbounded scenes}
\name{Chenhao Zhang$^{\ast}$,  Yezhi Shen$^{\ast }$,  Fengqing Zhu}
\address{Elmore Family School of Electrical and Computer Engineering \\Purdue University, West Lafayette, Indiana, U.S.A.
}

\begin{document}
\twocolumn[{
\renewcommand\twocolumn[1][]{#1}%
\maketitle

\thispagestyle{empty}
\begin{center}
    \centering
    \captionsetup{type=figure}
    \vspace{-0.5cm}
    \includegraphics[width=.95\textwidth]{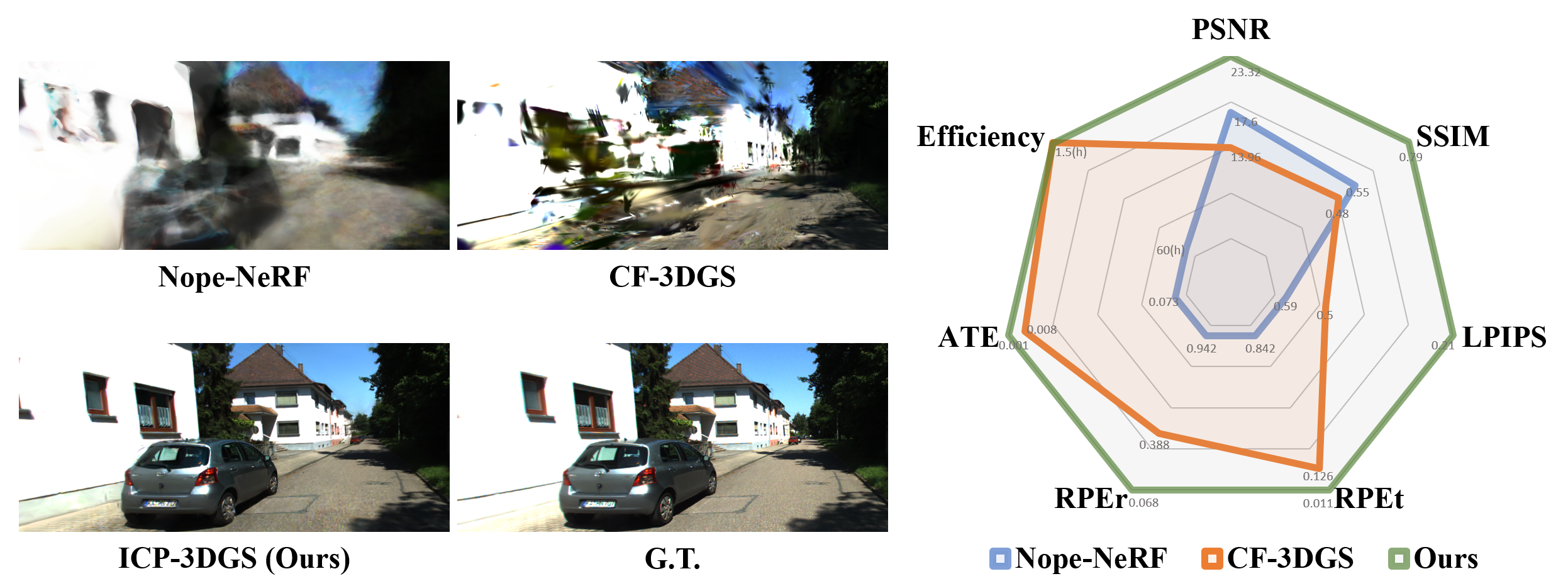}
    \captionof{figure}{Qualitative and quantitative comparisons in radar map of our proposed method vs. Nope-NeRF and CF-3DGS on outdoor unbounded scene reconstruction. Details of the quantitative comparison can be found in Section \ref{sec:Experiment}.}
\end{center}
}]

%
%

\begin{abstract}
In recent years, neural rendering methods such as NeRFs and 3D Gaussian Splatting (3DGS) have made significant progress in scene reconstruction and novel view synthesis. However, they heavily rely on preprocessed camera poses and 3D structural priors from structure-from-motion (SfM), which are challenging to obtain in out door scenarios. To address this challenge, we propose to incorporate Iterative Closest Point (ICP) with optimization-based refinement to achieve accurate camera pose estimation under large camera movements. Additionally, we introduce a voxel-based scene densification approach to guide the reconstruction in large-scale scenes. Experiments demonstrate that our approach ICP-3DGS outperforms existing methods in both camera pose estimation and novel view synthesis across indoor and outdoor scenes of various scales. Source code is available at \href{https://github.com/Chenhao-Z/ICP-3DGS}{https://github.com/Chenhao-Z/ICP-3DGS}.
\end{abstract}

\input{Sections/1_intro}
\input{Sections/3_method}
\input{Sections/4_experiment}
\input{Sections/5_conclusion}

\newpage
\newpage

\bibliographystyle{IEEEbib}

\bibliography{ref}

\end{document}

%% file: Sections/1_intro.tex
\section{Introduction}
\renewcommand{\thefootnote}{\fnsymbol{footnote}}
\footnotetext[1]{These authors contributed equally.}

\label{sec:intro}



\begin{figure*}[ht!]
    \centering
    \vspace{-0.3cm}
    \includegraphics[width=0.95\textwidth]{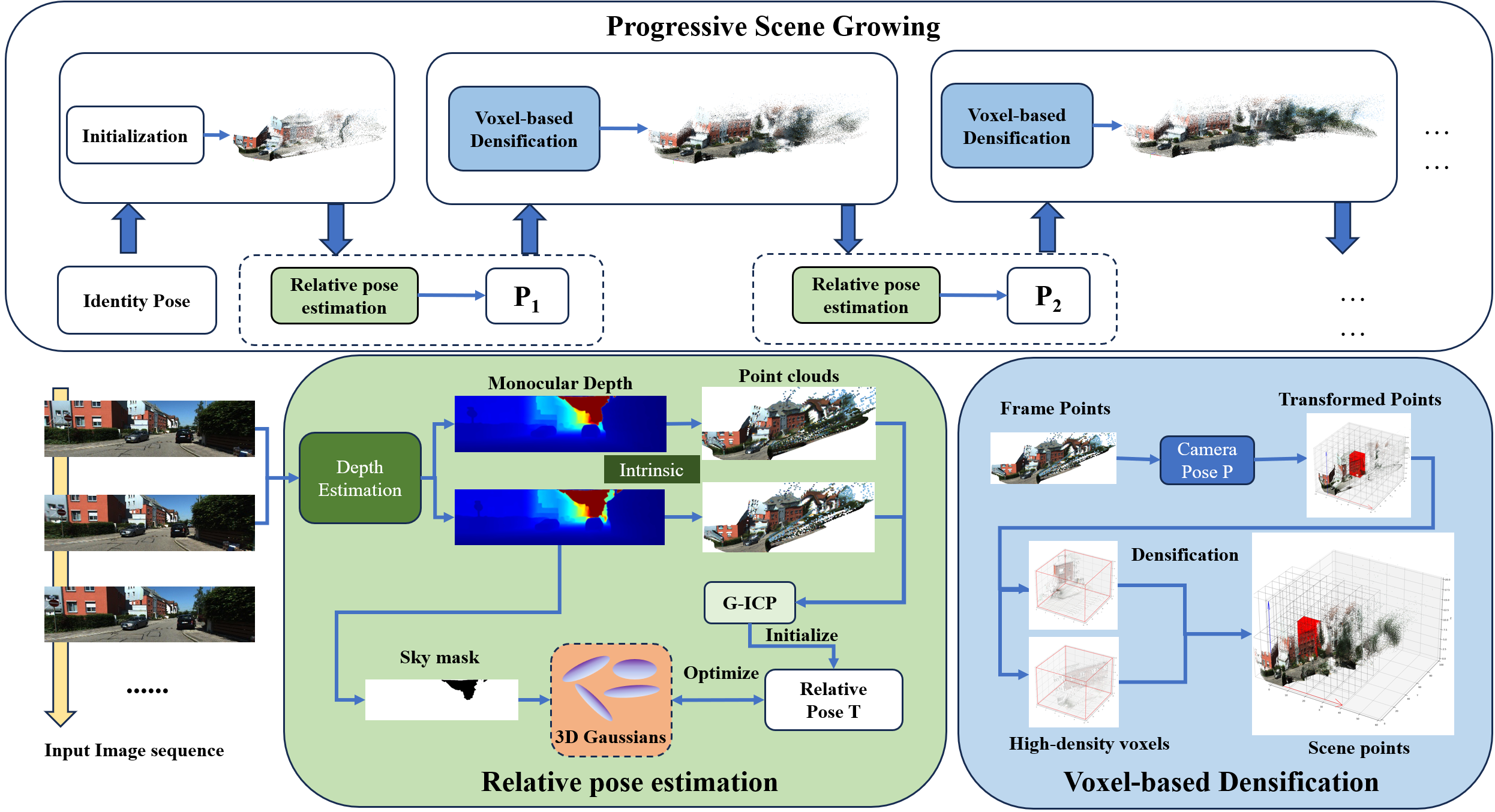}
    \vspace{-0.2cm}
    \caption{The overall framework of the proposed method consists of relative camera pose estimation and voxel-based densification.}
    \label{fig: overall}
    \vspace{-0.4cm}
\end{figure*}

Neural rendering has witnessed significant progress recently, with models like Neural Radiance Fields (NeRF) \cite{Nerf} and 3D Gaussian Splatting (3DGS) \cite{3DGS} emerging as leading techniques for photorealistic view synthesis and 3D reconstruction. 

Despite their success, NeRF and 3DGS share a critical limitation: they rely heavily on accurate camera poses and prior knowledge of 3D geometry for initialization. In practice, however, acquiring precise camera poses and reliable 3D priors are challenging. Structure-from-motion (SfM) \cite{SFM} , a commonly used substitute, also suffers several inherent drawbacks: it is computationally expensive, sensitive to feature extraction and matching errors, and prone to failure in scenes with sparse textures or large camera motions.

Recent SfM-free neural scene reconstruction methods jointly optimize neural representations and camera poses. NeRF\textminus \textminus \cite{Nerfmm} pioneered pose embedding co-optimization for NeRFs, with BarF \cite{Barf} and GarF \cite{Garf} later addressing gradient inconsistencies via coarse-to-fine encoding. However, these require good pose initialization. Nope-NeRF \cite{Nope-nerf} introduces a depth-based prior but suffers from long training times and suboptimal rendering. In 3D Gaussian Splatting, CF-3DGS \cite{CF-3DGS} refines affine-transformed Gaussians for pose estimation, while Free-SurGS \cite{FreeSurGS} employs optical flow loss, and IncEventGS \cite{inceventgs} integrates SLAM and event cameras for robustness. 

Despite promising results on bounded scenes where a single image can cover most of the reconstruction space, these methods struggle in large-scale, outdoor scenes, such as urban street environments. Such unbounded outdoor scenes are characterized by diverse object textures, ambiguous 3D geometry caused by frequent occlusions, and most importantly, vast spatial extents, which introduce new regions in every frame \cite{6img_to_3D}. Additionally, camera trajectories in these scenes often cover long distances with significant motion, making pose estimation even more challenging. 

In this paper, we propose the SfM-free 3DGS framework, consisting of effective camera pose estimation and adaptive scene growth for both bounded and unbounded scenes. The camera pose estimation module is constructed using a monocular depth estimation model with the Generalized Iterative Closest Point (G-ICP) algorithm \cite{G-ICP} to obtain a coarse estimation of the relative pose between frames as initialization. The scene growth module leverages occupancy information from the estimated depth map to adaptively densify the scene.

Our contributions are as follows:
\begin{itemize}
    \item We introduce monocular depth estimation with G-ICP to guide camera pose estimation, effectively addressing the challenges of pose estimation under large camera movements.
    \item We propose a voxel density-based 3DGS densification method to assist the initialization of 3DGS without SfM preprocessing, making it particularly suitable for large-scale scene reconstruction.
    \item Overall, the proposed Sfm free 3DGS approach significantly outperforms existing pose-unknown 3DGS methods on large-scale scenes by more than 9 dB PSNR.
\end{itemize}

%% file: Sections/3_method.tex
\section{METHODOLOGY}
\label{sec:method}

\subsection{Preliminary}
\textbf{Problem formulation.} Given a sequence of $N$ unposed RGB images frames, we denote it as $\{I_i\}^{N}_{i=1}$, where $I_{i} \in  \mathbb{R}^{H \times W \times 3}$ and K is the camera intrinsics. Our objective is to recover the camera pose $P_i$ for each of the images $I_i$ and reconstruct a 3D Gaussian splatting (3DGS) representation $G$ for the whole scene simultaneously. The output is the synthesis of photorealistic rendering $\hat{I}s$ from novel viewpoints.

\noindent\textbf{3D Gaussian Splatting.} 3D Gaussian Splatting (3DGS) \cite{3DGS} is an explicit point-based 3D scene representation, which represents a 3D scene into a group of oriented Gaussian ellipsoids. Each Gaussian ellipsoid is characterized by the 3D center point $\mu$ and a covariance matrix $\Sigma$:
\begin{equation}
    G(x) = e^{-\frac{1}{2}(x-\mu)^T\Sigma ^{-1}(x-\mu)}
\end{equation}
The color of the Gaussian ellipsoid is parameterized by a spherical harmonics (SH) coefficients vector $c$ and opacity $\alpha$.  Given a fixed camera view, 3DGS renders by approximating the projection of all the Gaussian ellipsoids along the depth dimension into the image pixel coordinate. With the camera view defined by the view transform matrix $W$ (also known as the camera pose), the formerly mentioned rendering process can be expressed as: 
\begin{equation}
    \Sigma^{2D} = JW\Sigma W^TJ^T
\end{equation} , where $\Sigma^{2D}$ is the covariance matrix in the pixel plane and $J$ is the Jacobian of the affine approximation of the projective transformation. The color is rendered as the weighted sum of the colors of sorted Gaussian ellipsoids overlapping at that pixel.

\subsection{Camera pose estimation} 
If we set the camera pose of the first frame as the identity pose matrix: $P_1 = [\mathbb{I}|O]$, the camera pose of the $i$th frame $P_i$ can be calculated as the relative camera pose with the first frame as the reference frame. However, directly estimating each frame's camera pose with respect to the first frame is very challenging. An alternative solution is to estimate the relative camera pose between every two adjacent frames. For all temporally adjacent frame pairs $(I_t,I_{t+1})$ in the input frames, we estimate their relative pose $R_{t,t+1}$. Then the camera pose of the $i$-th frame $P_i$ can be expressed as: 
\begin{equation}
    P_i = \prod_t^i R_{t,t+1}
\end{equation}

Following the approach outlined in CF-3DGS \cite{CF-3DGS}, estimating the relative camera pose is equivalent to estimating the transformation of a set of 3D Gaussian points. Given a frame pair $(I_t,I_{t+1})$, we first use a pre-trained monocular depth model, $i.e.$, Metric3DV2 \cite{Metric3DV2} to generate the depth map $D_t$ of the previous frame $I_t$. Then we use the identity camera pose $[\mathbb{I}|O]$ and the camera intrinsic matrix $K$ to lift the depth map into 3D points and set it as the initialization of 3DGS. By minimizing the photometric loss between the rendered image and the previous frame $I_t$, we can get a set of 3D Gaussians $G_{t}$ that is overfitted on this single frame.

\begin{equation}
    G_t = \arg\!\min\limits_{R_t,S_t,c_t,\alpha_t} \mathcal{L}_{rgb}(\mathcal{R}(G_t),I_t)
\end{equation}
where, $\mathcal{R} $ stands for the rendering process and the photometric loss $\mathcal{L}_{rgb}$ is a weighted sum of the $L_1$ loss and the D-SSIM loss:
\begin{equation}
    \mathcal{L}_{rgb} = (1-\lambda)\mathcal{L}_1 + \lambda \mathcal{L}_{D-SSIM}
\end{equation}

In order to estimate the relative camera pose $R_{t,t+1}$, we apply a learnable SE-3 affine transformation $T_{t,t+1}$ to $G_t$ \cite{Lie_group} to transform the previous frame $I_t$ into the current frame $I_{t+1}$, which is supervised by minimizing the rendering loss between the rendered image and the current frame $I_{t+1}$. Different from CF-3DGS\cite{CF-3DGS}, we use just $L_1$ loss in this optimization process because we find that the full photometric loss will bring significant instability:
\begin{equation}
    T_{t,t+1} = \arg\min\limits_{T_{t,t+1}} \mathcal{L}_{1}(\mathcal{R}(T_t \odot G_t),I_{t+1})
\end{equation}

\noindent\textbf{G-ICP guiding.} The optimization method above assumes that the relative camera motion between adjacent frames is small, which works well in scenarios where there is sufficient overlap between frames. However, in real-world applications, especially in street scenes, large camera motions between frames are common, and the overlap between consecutive frames are small, leading to difficulties in the optimization process.
To address this challenge, we leverage the Iterative Closest Point (ICP) \cite{ICP} algorithm, which is widely used in point cloud registering. G-ICP \cite{G-ICP} uses probabilistic models to generalize ICP and improves robustness.

For a frame pair $(I_t,I_{t+1})$, we lift both frames into 3D points and get two corresponding point cloud $(Pcd_t,Pcd_{t+1})$. We first use the G-ICP to obtain a transformation $\hat{T}_{t,t+1}$ that minimize the Euclidean distance between $Pcd_t$ and $Pcd_{t+1}$'s corresponding points. We then set it as an initialization for the aforementioned camera pose optimization process. While this initial estimate may not be highly accurate due to the depth error, it provides a reasonable starting point to guide further optimization. In our experiments, we observe that such an initialization significantly alleviates the problem of direct optimization tends to get stuck in local minima when the camera motion is too large.

\noindent\textbf{Sky Removal.} Additionally, the true depth of the sky is at infinity in outdoor scenes, which is challenging in 3D structure modeling. The monocular depth models usually choose to clamp the depth into a pre-defined value range, making the depth of this part unreliable. When the sky is included in camera pose optimization, it can lead to significant errors. To mitigate this issue, we propose to use a sky mask to exclude sky regions from the relative camera pose optimization. By leveraging the maximum depth value from the depth map $D_t$, we can define the sky mask $M_t$ without additional pre-trained semantic segmentation models:
\begin{equation}
   M_t = \mathbb{I}(D_t = \max(D_t))
\end{equation}
By applying such masks to rendering results, we ensure that the optimization focuses on regions with accurate depth information.

\subsection{Progressive scene growing} 
Next, we use the estimated camera poses to reconstruct the scene. Methods such as CF-3DGS \cite{CF-3DGS} and other works \cite{FreeSurGS} simply initialize Gaussians for the entire scene using the point cloud from the first frame and then iteratively expand it frame-by-frame through the standard densification process of 3DGS. However, this approach has clear limitations. The point cloud from the first frame often fails to cover the entire scene, especially in large-scale scenarios. The densification process in 3DGS relies on splitting and replicating Gaussians based on the gradient of each Gaussian. In areas without any initial points, this process struggles to generate new Gaussians \cite{Revise-Densification}, leading to suboptimal rendering in these regions. A straightforward approach is to integrate the new frame's point cloud into the scene's Gaussians, but this causes significant overlap and unnecessary points. Overlapping erroneous points can disrupt gradient computation, hinder Gaussian pruning, and degrade rendering quality. To address the aforementioned issues, we propose a voxel-based densification approach.

Using the estimated camera pose $ P_t$, we transform the projected 3D points of the current frame $Pcd_t$ into the scene coordinates: 
\begin{equation}
    \hat{Pcd}_t = P_t\odot Pcd_t
\end{equation}
We then compute the bounding of the transformed points $\hat{Pcd}_t$, donated as $B_t$. We are able to find the centers of the points within $B_t$ in the current Gaussian, donated as $g_t$. By voxelizing both $\hat{Pcd}_t$ and $g_t$ using a predetermined voxel size, we are enabled to compare the density of points in each corresponding voxel pairs. 

When the density ratio of a voxel exceeds the threshold, it indicates that this region requires additional points due to the lack of coverage from the existing scene representation. We then add the points from the $\hat{Pcd}_t$ within this voxel to the scene. Conversely, if the ratio is below the threshold, the region is considered adequately covered and no points are needed to be added.
This density-driven, voxel-based approach effectively identifies regions of the scene that need rapid expansion, thus better guiding the densification process on unbounded scenes. 

%% file: Sections/4_experiment.tex
\section{Experiment} 
\label{sec:Experiment}

\subsection{Experimental Setting}
\begin{table}[h!]
    \centering
    \renewcommand{\arraystretch}{1.3} 
    \setlength{\tabcolsep}{4pt} 
    \resizebox{\columnwidth}{!}{ 
    \begin{tabular}{lccc}
        \toprule
         & \textbf{CF-3DGS} & \textbf{Nope-NeRF} & \textbf{Ours} \\
         & PSNR $\uparrow$ / SSIM $\uparrow$ / LPIPS $\downarrow$ & 
           PSNR $\uparrow$ / SSIM $\uparrow$ / LPIPS $\downarrow$ & 
           PSNR $\uparrow$ / SSIM $\uparrow$ / LPIPS $\downarrow$ \\
       \midrule
        Seq 1 & 13.59 / 0.59 / 0.45 & 17.01 / 0.65 / 0.53 & \textbf{23.77} / \textbf{0.84} / \textbf{0.19} \\
        Seq 2 & 14.85 / 0.48 / 0.49 & 16.69 / 0.50 / 0.63 & \textbf{22.81} / \textbf{0.78} / \textbf{0.22} \\
        Seq 3 & 13.52 / 0.40 / 0.56 & 18.02 / 0.50 / 0.62 & \textbf{23.76} / \textbf{0.77} / \textbf{0.21} \\
        Seq 4 & 12.14 / 0.41 / 0.54 & 16.92 / 0.51 / 0.61 & \textbf{21.49} / \textbf{0.74} / \textbf{0.23} \\
        Seq 5 & 15.72 / 0.50 / 0.46 & 19.37 / 0.57 / 0.58 & \textbf{24.78} / \textbf{0.80} / \textbf{0.20} \\
        \midrule
        AVG  & 13.96 / 0.48 / 0.50 & 17.60 / 0.55 / 0.59 & \textbf{23.32} / \textbf{0.79} / \textbf{0.21} \\
        \bottomrule
    \end{tabular}
    }
    \caption{Novel view synthesis results on KITTI-360. Best results are in \textbf{bold}.}
    \vspace{-0.7cm}
    \label{tab:NVS_KITTI}
\end{table}

\begin{table}[h!]
    \centering
    \renewcommand{\arraystretch}{1.5} 
    \vspace{0mm} 
    \resizebox{\columnwidth}{!}{ 
    \begin{tabular}{lccc}
        \toprule
         & \textbf{Nope-NeRF} & \textbf{CF-3DGS} & \textbf{Ours} \\
        & PSNR $\uparrow$ / SSIM $\uparrow$ / LPIPS $\downarrow$ 
        & PSNR $\uparrow$ / SSIM $\uparrow$ / LPIPS $\downarrow$ 
        & PSNR $\uparrow$ / SSIM $\uparrow$ / LPIPS $\downarrow$ \\
        \midrule
        Church   & 25.17 / 0.73 / 0.39 & 30.23 / \textbf{0.93} / 0.11 & \textbf{30.54 / 0.93 / 0.09} \\
        Barn     & 26.35 / 0.69 / 0.44 & 31.23 / 0.90 / 0.10 & \textbf{34.15 / 0.95 / 0.06} \\
        Museum   & 26.77 / 0.76 / 0.35 & 29.91 / 0.91 / 0.11 & \textbf{31.91 / 0.94 / 0.07} \\
        Family   & 26.01 / 0.74 / 0.41 & 31.27 / 0.94 / 0.07 & \textbf{32.50 / 0.95 / 0.06} \\
        Horse    & 27.64 / 0.84 / 0.26 & 33.94 / \textbf{0.96} / 0.05 & \textbf{34.29 / 0.96 / 0.05} \\
        Ballroom & 25.33 / 0.72 / 0.38 & 32.47 / 0.96 / 0.07 & \textbf{34.64 / 0.97 / 0.03} \\
        Francis  & 29.48 / 0.80 / 0.38 & 32.72 / 0.91 / 0.14 & \textbf{33.76 / 0.93 / 0.11} \\
        Ignatius  & 23.96 / 0.61 / 0.47 & 28.43/ 0.90 / 0.09 & \textbf{29.02 / 0.94 / 0.08} \\
        \midrule
        AVG      & 26.34 / 0.74 / 0.39 & 31.28 / 0.93 / 0.09 & \textbf{32.60 / 0.94 / 0.07} \\
        \bottomrule
    \end{tabular}
    }
    \caption{Novel view synthesis results on Tanks and Temples. Best results are in \textbf{bold}.}
    \label{tab:NVS_TT}
\end{table}

\noindent\textbf{Datasets.} 
 
Our method is evaluated against other pose-unknown methods on KITTI-360 \cite{KITTI-360}, and Tanks and Temples \cite{Tanks&Temple} datasets. For KITTI-360 dataset, we follow \cite{Lightening-nerf} to select five sequences that mainly contain static traffic agents (i.e. parked cars). The KITTI-360 dataset is representative of large-scale outdoor scenes, in which each video sequence is recorded over 100 meters of displacement. The Tanks and Temples dataset include both indoor and outdoor scenes on a smaller scale of \~10 meters. For both datasets, we follow Nope-NeRF \cite{Nope-nerf} to construct the training and testing frames sets. For each scene in the datasets, one in every eight frames is selected for testing, with the remaining frames used for training, except for the `Family' scene in the Tanks and Temples dataset, where every other frame is selected for training and testing. 

\begin{table}[h!]
    \centering
    \renewcommand{\arraystretch}{1.3} 
    \setlength{\tabcolsep}{4pt} 
    \resizebox{\columnwidth}{!}{ 
    \begin{tabular}{lccc}
        \toprule
         & \textbf{Nope-NeRF} & \textbf{CF-3DGS} & \textbf{Ours} \\
         & RPEt $\downarrow$ / RPEr $\downarrow$ / ATE $\downarrow$ & 
           RPEt $\downarrow$ / RPEr $\downarrow$ / ATE $\downarrow$ & 
           RPEt $\downarrow$ / RPEr $\downarrow$ / ATE $\downarrow$ \\
       \midrule
        Seq 1 & 0.923 / 1.050 / 0.075 & 0.144 / 0.450 / 0.006 & \textbf{0.008} / \textbf{0.058} / \textbf{0.001} \\
       Seq 2 & 0.732 / 0.800 / 0.078 & 0.087 / 0.335 / 0.007 & \textbf{0.008} / \textbf{0.071} / \textbf{0.000} \\
        Seq 3 & 0.802 / 1.026 / 0.056 & 0.143 / 0.341 / 0.014 & \textbf{0.010} / \textbf{0.055} / \textbf{0.001} \\
        Seq 4 & 1.121 / 0.948 / 0.075 & 0.117 / 0.482 / 0.006 & \textbf{0.015} / \textbf{0.086} / \textbf{0.001} \\
        Seq 5 & 0.631 / 0.886 / 0.082 & 0.139 / 0.331 / 0.006 & \textbf{0.013} / \textbf{0.068} / \textbf{0.001} \\
        \midrule
       AVG    & 0.842 / 0.942 / 0.073 & 0.126 / 0.388 / 0.008 & \textbf{0.011} / \textbf{0.068} / \textbf{0.001} \\
        \bottomrule
    \end{tabular}
    }
    \caption{Camera pose estimation results on KITTI-360. Best results are in \textbf{bold}.}
    \label{tab:CE_KITTI}
\end{table}

\begin{table}[h!]
    \centering
    \renewcommand{\arraystretch}{1.3} 
    \setlength{\tabcolsep}{4pt} 
    \resizebox{\columnwidth}{!}{ 
    \begin{tabular}{lccc}
        \toprule
         & \textbf{Nope-NeRF} & \textbf{CF-3DGS} & \textbf{Ours} \\
       \midrule
        KITTI-360 & $\sim$ 60h & $\sim$ 1h 15min & $\sim$ 1h 30min \\
       Tanks and Temples &$\sim$ 30h &$\sim$ 1h 30min& $\sim$ 1h 30min \\
        \bottomrule
    \end{tabular}
    }
    \caption{Training time comparison. }
    \label{tab:Time_KITTI}
\end{table}

\begin{figure*}[t]
    \centering
    \begin{minipage}{\linewidth}
        \centering
        \subfloat[Nope-NeRF]{\includegraphics[width=0.49\linewidth]{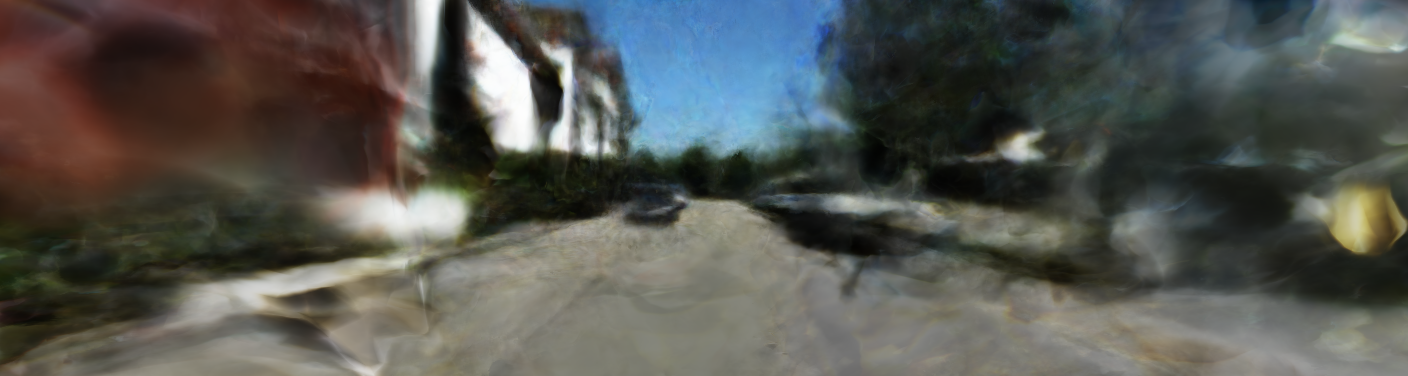}}
        \subfloat[CF-3DGS]{\includegraphics[width=0.49\linewidth]{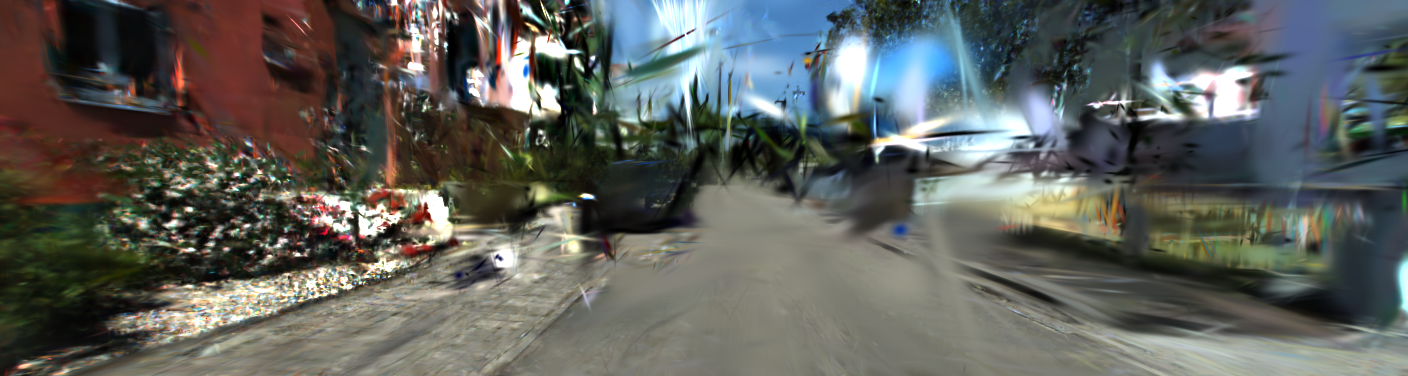}} \\
        \subfloat[Ours]{\includegraphics[width=0.49\linewidth]{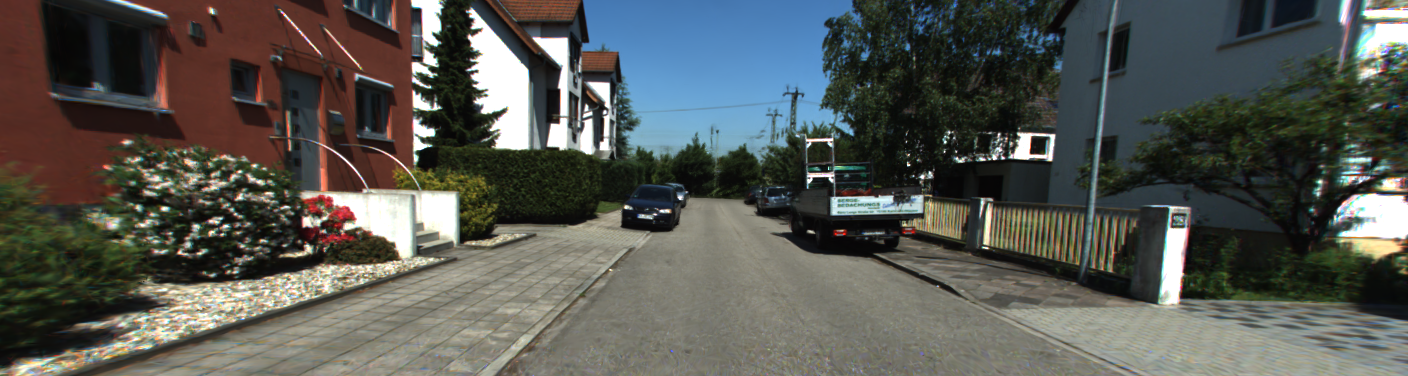}}
        \subfloat[Ground Truth]{\includegraphics[width=0.49\linewidth]{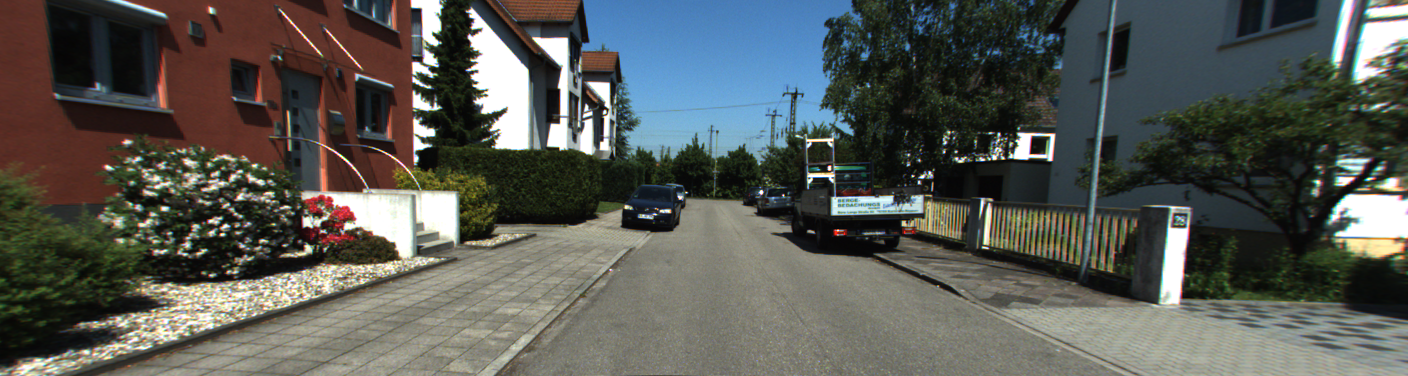}}
    \end{minipage}

\caption{Qualitative comparison for novel view synthesis on KITTI-360.}
    \label{fig:qualitative_results_kitti}
\end{figure*}

\begin{figure}[h!]
    \centering
    \subfloat[Nope-NeRF]{\includegraphics[width=0.48\linewidth]{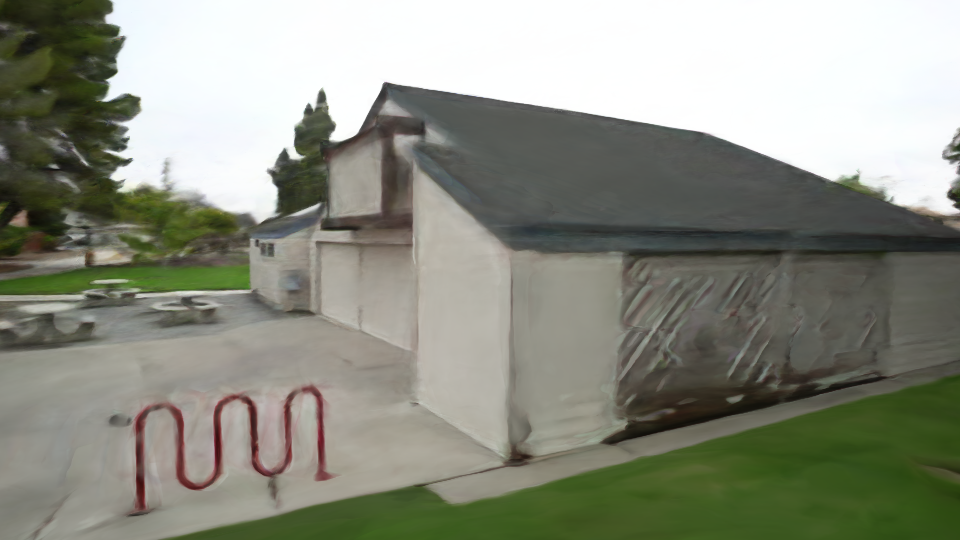}}
    \subfloat[CF-3DGS]{\includegraphics[width=0.48\linewidth]{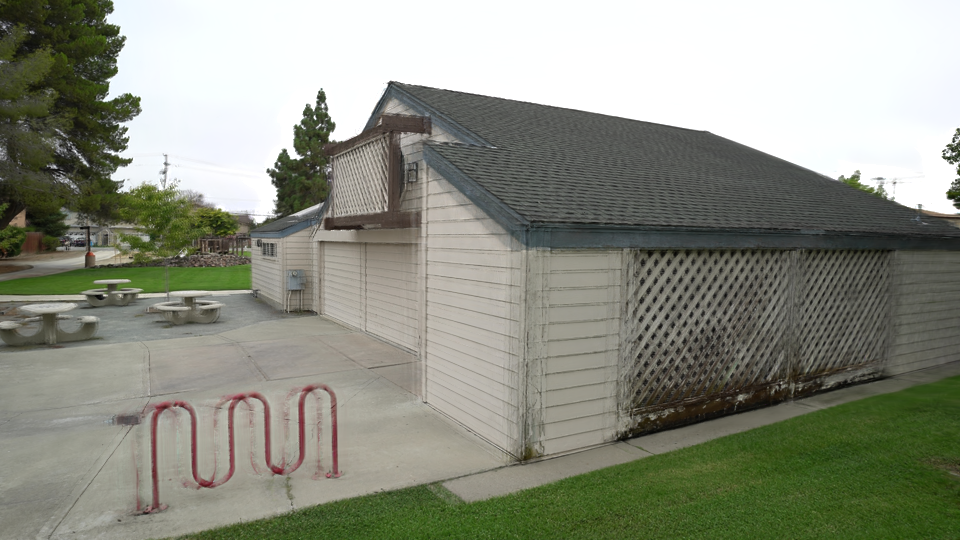}} \\
    \subfloat[Ours]{\includegraphics[width=0.48\linewidth]{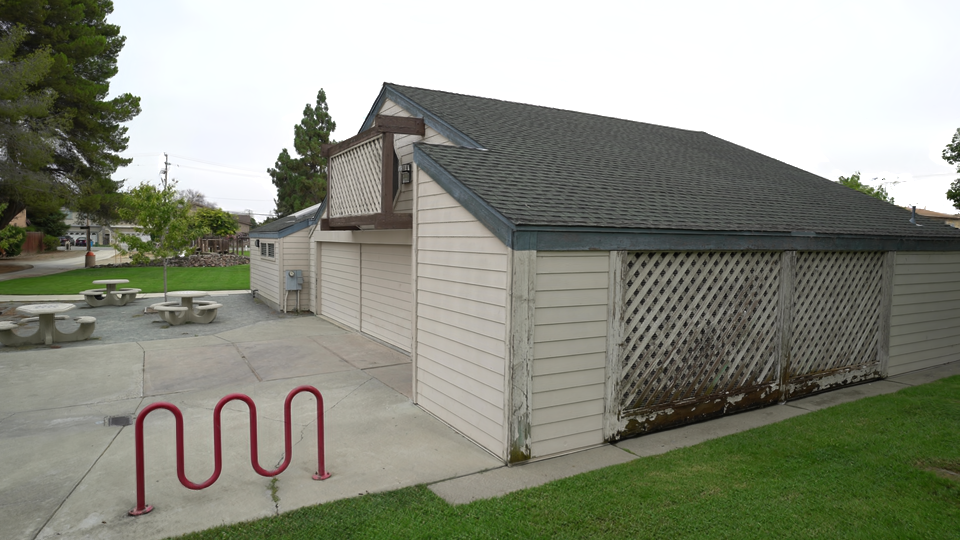}}
    \subfloat[Ground Truth]{\includegraphics[width=0.48\linewidth]{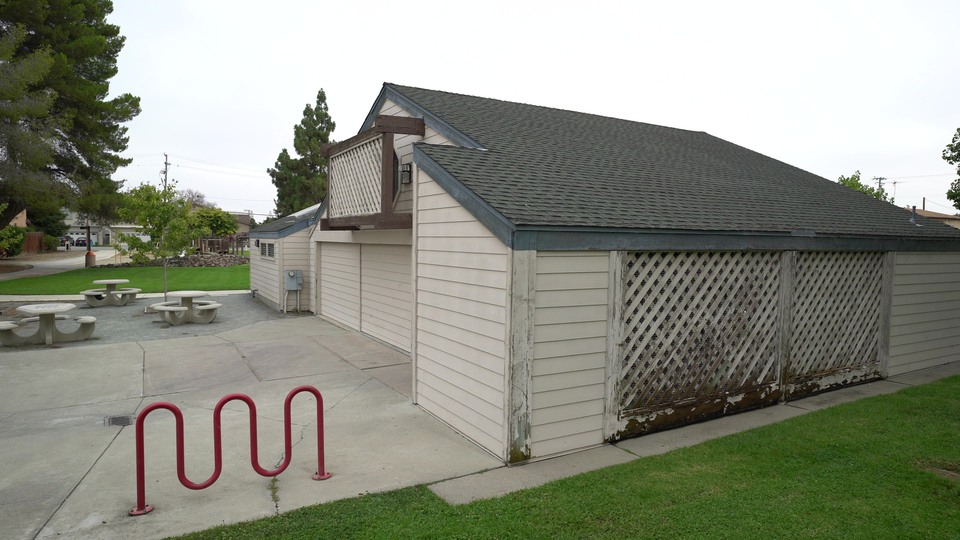}} 
    \caption{Qualitative comparison for novel view synthesis on Tanks and Temples.}
    \label{fig:qualitative_comparison}
    \vspace{-0.5cm}
\end{figure}

\noindent\textbf{Metrics.} For the novel view synthesis task, we use standard evaluation metrics, including PSNR, SSIM \cite{SSIM}, and LPIPS \cite{lpips}. For the camera pose estimation task, we use Absolute Trajectory Error (ATE), Relative Pose Error (RPE$_r$, RPE$_t$), following the settings in \cite{Nope-nerf, Barf}.

\subsection{Quantitative Comparison} 
We conduct quantitative evaluations against two existing leading pose-unknown method, Nope-NeRF \cite{Nope-nerf} and CF-3DGS \cite{CF-3DGS} on both novel view synthesis and camera pose estimation. Following the idea of NeRF\textminus \textminus \cite{Nerfmm}, we freeze our trained 3DGS model and minimize the photometric error between the synthesized images and the test views to obtain the testing camera poses. We apply this procedure to all methods during evaluation to ensure a fair comparison. 

The quantitative comparison result for novel view synthesis on KITTI-360 \cite{KITTI-360} is reported in Table \ref{tab:NVS_KITTI}, and the result on Tanks and Temples \cite{Tanks&Temple} is reported in Table \ref{tab:NVS_TT}. It can be seen that our method outperforms existing results on unbounded scene reconstruction tasks in every testing scene and we achieve a significant 5.7 dB PSNR improvement on average. It can also be observed that, even though our method is primarily designed for large-scale scenes, we still achieve over 1.3 dB improvement in PSNR compared to CF-3DGS and Nope-NeRF, on the bounded Tanks and Temples dataset. This clearly demonstrates the excellent generalization ability of our method.

Since Tanks and Temple dataset does not provide ground-truth camera poses, we only report the camera pose evaluation results on KITTI-360, as shown in Table \ref{tab:CE_KITTI}. To ensure a fair comparison, we align the scale and trajectory for Nope-NeRF and CF-3DGS before computing the camera pose error. Results show that our method surpasses existing methods in terms of camera trajectory estimation thoroughly.

The average training time results on both datasets are reported in Table \ref{tab:Time_KITTI}. Our method consistently outperforms the others across all metrics on both tasks, while only slightly increasing the training time compared to CF-3DGS, and remains much faster than Nope-NeRF.

\subsection{Qualitative Comparison} 
The qualitative results on KITTI-360 are shown in Figure \ref{fig:qualitative_results_kitti}, while the qualitative results on Tanks and Temples are presented in Figure 3. As illustrated in Figure \ref{fig:qualitative_results_kitti}, both baseline methods struggle on the large-scale unbounded dataset. The results from Nope-NeRF are overly blurred, making it difficult to discern details, whereas CF-3DGS exhibits noticeable artifacts. Only our method maintains high quality rendering performance.

\begin{table}[h]
    \vspace{-0.1cm}
    \centering
    \resizebox{\columnwidth}{!}{ 
    \begin{tabular}{c|c|c|c|ccc}
        \toprule
        \textbf{Base} & \textbf{S mask} & \textbf{G-ICP} & \textbf{VB-Den} & \textbf{PSNR ↑} & \textbf{SSIM ↑} & \textbf{LPIPS ↓} \\
        \midrule
        \checkmark &  &  &  &  16.23 & 0.52 & 0.47 \\
        \checkmark &  &  & \checkmark & 16.57 & 0.55 & 0.40 \\
        \checkmark & \checkmark &  &  & 17.39 & 0.61 & 0.42 \\
        \checkmark & \checkmark & \checkmark &  & 20.89 & 0.67 & 0.38 \\
        \checkmark & \checkmark & \checkmark & \checkmark & 23.32 & 0.79 & 0.21 \\
        \bottomrule
    \end{tabular}}
    \caption{Ablation study on KITTI-360.}
    \vspace{-0.7cm}
    \label{tab:Ablation_scene}
\end{table}

\subsection{Ablation Study.} 
We conduct an ablation study on KITTI-360, reporting average PSNR, SSIM, and LPIPS in Table \ref{tab:Ablation_scene}. “S mask” denotes the sky mask, and “VB-Den” represents voxel-based densification. The base setup follows CF-3DGS [8] but replaces the monocular depth model with Metric3Dv2 [13]. Voxel-based densification alone has minimal impact, but when combined with our auxiliary pose estimation module, it greatly improves rendering quality, highlighting the importance of accurate camera poses for 3D point initialization and validating the effectiveness of our modules.

%% file: Sections/5_conclusion.tex
\section{Conclusion.}
\label{sec:Conclusion}
We propose an SFM-free 3D Gaussian method for large-scale unbounded scenes. To address the challenges of large camera movements and inconsistent depth in outdoor scenes, we introduce a sky mask to separate the background and utilize G-ICP in a two-step optimization process to achieve robust camera pose estimation. Furthermore, we propose a voxel-based scene expansion strategy to guide the Gaussian ellipsoids' growth in areas that lack initial points. Experiments show that our method outperforms existing SFM-free approaches in both camera pose estimation and novel view synthesis tasks and demonstrates strong generalization across different scene scales, whether indoor or outdoor.

%% file: main.bbl
\begin{thebibliography}{10}

\bibitem{Nerf}
B.~Mildenhall, P.~P. Srinivasan, M.~Tancik, J.~T. Barron, R.~Ramamoorthi, and R.~Ng,
\newblock ``Nerf: representing scenes as neural radiance fields for view synthesis,''
\newblock {\em Commun. ACM}, vol. 65, no. 1, pp. 99–106, Dec. 2021.

\bibitem{3DGS}
B.~Kerbl, G.~Kopanas, T.~Leimkuehler, and G.~Drettakis,
\newblock ``3d gaussian splatting for real-time radiance field rendering,''
\newblock {\em ACM Trans. Graph.}, vol. 42, no. 4, July 2023.

\bibitem{SFM}
J.~L. Schönberger and J.-M. Frahm,
\newblock ``Structure-from-motion revisited,''
\newblock {\em 2016 IEEE Conference on Computer Vision and Pattern Recognition (CVPR)}, pp. 4104--4113, 2016.

\bibitem{Nerfmm}
Z.~Wang, S.~Wu, W.~Xie, M.~Chen, and V.~A. Prisacariu,
\newblock ``Nerf--: Neural radiance fields without known camera parameters,''
\newblock {\em arXiv preprint}, vol. arXiv:2102.07064, 2022.

\bibitem{Barf}
C.-H. Lin, W.-C. Ma, A.~Torralba, and S.~Lucey,
\newblock ``Barf: Bundle-adjusting neural radiance fields,''
\newblock {\em 2021 IEEE/CVF International Conference on Computer Vision (ICCV)}, pp. 5721--5731, 2021.

\bibitem{Garf}
S.-F. Chng, S.~Ramasinghe, J.~Sherrah, and S.~Lucey,
\newblock ``Gaussian activated neural radiance fields for high fidelity reconstruction and pose estimation,''
\newblock {\em Computer Vision – ECCV 2022: 17th European Conference, Tel Aviv, Israel, October 23–27, 2022, Proceedings, Part XXXIII}, p. 264–280, 2022.

\bibitem{Nope-nerf}
W.~Bian, Z.~Wang, K.~Li, and J.-W. Bian,
\newblock ``Nope-nerf: Optimising neural radiance field with no pose prior,''
\newblock {\em 2023 IEEE/CVF Conference on Computer Vision and Pattern Recognition (CVPR)}, pp. 4160--4169, 2023.

\bibitem{CF-3DGS}
Y.~Fu, X.~Wang, S.~Liu, A.~Kulkarni, J.~Kautz, and A.~A. Efros,
\newblock ``Colmap-free 3d gaussian splatting,''
\newblock {\em 2024 IEEE/CVF Conference on Computer Vision and Pattern Recognition (CVPR)}, pp. 20796--20805, 2024.

\bibitem{FreeSurGS}
J.~Guo, J.~Wang, D.~Kang, W.~Dong, W.~Wang, and Y.-h. Liu,
\newblock ``{ Free-SurGS: SfM-Free 3D Gaussian Splatting for Surgical Scene Reconstruction },''
\newblock {\em proceedings of Medical Image Computing and Computer Assisted Intervention -- MICCAI 2024}, vol. LNCS 15007, October 2024.

\bibitem{inceventgs}
J.~Huang, C.~Dong, and P.~Liu,
\newblock ``Inceventgs: Pose-free gaussian splatting from a single event camera,''
\newblock {\em arXiv preprint}, vol. arXiv:2410.08107, 2024.

\bibitem{6img_to_3D}
T.~Gieruc, M.~Kästingschäfer, S.~Bernhard, and M.~Salzmann,
\newblock ``6img-to-3d: Few-image large-scale outdoor driving scene reconstruction,''
\newblock {\em arXiv preprint}, vol. arXiv:2404.12378, 2024.

\bibitem{G-ICP}
A.~Segal, D.~Hähnel, and S.~Thrun,
\newblock ``Generalized-icp.,''
\newblock {\em Robotics: Science and Systems}, vol. 2, pp. 435, 2009.

\bibitem{Metric3DV2}
M.~Hu, W.~Yin, C.~Zhang, Z.~Cai, X.~Long, H.~Chen, K.~Wang, G.~Yu, C.~Shen, and S.~Shen,
\newblock ``Metric3d v2: A versatile monocular geometric foundation model for zero-shot metric depth and surface normal estimation,''
\newblock {\em IEEE Transactions on Pattern Analysis and Machine Intelligence}, vol. 46, no. 12, pp. 10579–10596, December 2024.

\bibitem{Lie_group}
Z.~Teed and J.~Deng,
\newblock ``Tangent space backpropagation for 3d transformation groups,''
\newblock {\em 2021 IEEE/CVF Conference on Computer Vision and Pattern Recognition (CVPR)}, pp. 10333--10342, 2021.

\bibitem{ICP}
P.~Besl and N.~D. McKay,
\newblock ``A method for registration of 3-d shapes,''
\newblock {\em IEEE Transactions on Pattern Analysis and Machine Intelligence}, vol. 14, no. 2, pp. 239--256, 1992.

\bibitem{Revise-Densification}
S.~Rota~Bul\`{o}, L.~Porzi, and P.~Kontschieder,
\newblock ``Revising densification in gaussian splatting,''
\newblock {\em Computer Vision – ECCV 2024: 18th European Conference, Milan, Italy, September 29–October 4, 2024, Proceedings, Part LXIII}, p. 347–362, 2024,
\newblock Milan, Italy.

\bibitem{KITTI-360}
Y.~Liao, J.~Xie, and A.~Geiger,
\newblock ``Kitti-360: A novel dataset and benchmarks for urban scene understanding in 2d and 3d,''
\newblock {\em IEEE Transactions on Pattern Analysis and Machine Intelligence}, vol. 45, no. 3, pp. 3292--3310, 2023.

\bibitem{Tanks&Temple}
A.~Knapitsch, J.~Park, Q.-Y. Zhou, and V.~Koltun,
\newblock ``Tanks and temples: benchmarking large-scale scene reconstruction,''
\newblock {\em ACM Trans. Graph.}, vol. 36, no. 4, July 2017.

\bibitem{Lightening-nerf}
J.~Cao, Z.~Li, N.~Wang, and C.~Ma,
\newblock ``Lightning nerf: Efficient hybrid scene representation for autonomous driving,''
\newblock {\em 2024 IEEE International Conference on Robotics and Automation (ICRA)}, pp. 16803--16809, 2024.

\bibitem{SSIM}
Z.~Wang, A.~Bovik, H.~Sheikh, and E.~Simoncelli,
\newblock ``Image quality assessment: from error visibility to structural similarity,''
\newblock {\em IEEE Transactions on Image Processing}, vol. 13, no. 4, pp. 600--612, 2004.

\bibitem{lpips}
R.~Zhang, P.~Isola, A.~A. Efros, E.~Shechtman, and O.~Wang,
\newblock ``The unreasonable effectiveness of deep features as a perceptual metric,''
\newblock {\em 2018 IEEE/CVF Conference on Computer Vision and Pattern Recognition}, pp. 586--595, 2018.

\end{thebibliography}
